\documentclass[12pt]{article}
\usepackage{amsmath}
\usepackage{amssymb}
\usepackage{epsfig}
\usepackage{cite}
\usepackage{color,colordvi}

\begin{document}
\vspace{0.1cm}
\begin{center}
{\Large\bf  Corpuscular Breaking of Supersymmetry}

\end{center}

\vspace{0.15 cm}

\begin{center}

{\bf Gia Dvali}$^{a,b,c}$ and {\bf Cesar Gomez}$^{a,e}$\footnote{cesar.gomez@uam.es}

\vspace{.6truecm}

{\em $^a$Arnold Sommerfeld Center for Theoretical Physics\\
Department f\"ur Physik, Ludwig-Maximilians-Universit\"at M\"unchen\\
Theresienstr.~37, 80333 M\"unchen, Germany}

{\em $^b$Max-Planck-Institut f\"ur Physik\\
F\"ohringer Ring 6, 80805 M\"unchen, Germany}

{\em $^c$Center for Cosmology and Particle Physics\\
Department of Physics, New York University\\
4 Washington Place, New York, NY 10003, USA}

{\em $^e$
Instituto de F\'{\i}sica Te\'orica UAM-CSIC, C-XVI \\
Universidad Aut\'onoma de Madrid,
Cantoblanco, 28049 Madrid, Spain}\\

\end{center}

\begin{abstract}
\noindent  
 
{\small 
Are topological solitons elementary or composites? We answer this question by drawing up a corpuscular formalism in which solitons are coherent states of quantum constituents. This naturally leads  to  a functional integral representation, in which the classical saddle point is reached as the most probable distribution of corpuscles in the $\hbar = 0$ limit and where quantum corpuscular corrections correspond to excursions away from such a distribution that occur only for finite $\hbar$.  Several striking features come up.  Topological charge emerges as a collective flow of quantum numbers carried  by individual corpuscles. Moreover, the corpuscular corrections are not reducible  to any known form of quantum corrections, such as loop expansions in the coupling constant  $\hbar g^2$ or semiclassical $e^{-1/\hbar g^2}$ effects. Corpuscular corrections are stronger and appear already at order $\sqrt{\hbar g^2}$. In SUSY theories quantum corpuscular corrections generically break supersymmetry.  We show that a domain wall which perturbatively  is a BPS state, violates all supersymmetries when the corpuscular effects are taken into account. The extension of the corpuscular structure to $D$-branes can lead to a built-in supersymmetry breaking mechanism in string theory, insensitive to technicalities such as moduli stabilization, with the SUSY breaking scale set by the string coupling times the $D$-brane tension.

 }

\end{abstract}

\thispagestyle{empty}
\clearpage

\section{Introduction}
Topologically stable field configurations, in short solitons, are the key tool of nonperturbative quantum field theory. These solitons are classically defined by a saddle point field configuration of the classical action taking boundary values in the vacuum manifold. The asymptotic behavior and the topology of the vacuum manifold account for the topological stability (topological charge) of the soliton. In addition to the topological charge the soliton is generically characterized by the tension and by a typical length, $L$,  representing the size of the region of space where most of the energy is localized. Once we move into quantum field theory we should think about these solitons as quantum states $|sol\rangle$. On the other hand the particle content of the theory is defined relative to a Poincare invariant vacuum state $|0\rangle$. These particles -- in terms of which we define the asymptotic S-matrix states -- have {\it zero topological charge} and therefore they are orthogonal to the soliton state $|sol\rangle$. This state of affairs means that the soliton as a quantum system cannot be understood as being composed out of the same type of particles that we use to define the asymptotic S-matrix states. But then, should we conclude that the soliton as a quantum system must be thought as {\it fundamental}? 

   In this note we shall answer this question negatively and explore the consequences.  
 We have suggested in \cite{GiaCearN} that solitons and black holes must have a corpuscular structure and have outlined some similarities among the two entities.  In this note we shall develop the corpuscular approach to solitons.  
 
 We would like to bring across the following  points:
 \begin{itemize}
  \item The non-perturbative objects such as solitons can be viewed as composite entities in certain well-defined sense of quantum constituency.  
 
 \item The corpuscular structure of solitons gives rise to new  corrections that are not accounted by  
standard effects, such as,  perturbative loop expansions in $\hbar g^2$ and  non-perturbative 
$e^{-{1 \over \hbar g^2}}$ type corrections.  Corpuscular effects can be much stronger  and generically are  non-analytic in  $\hbar g^2$.   In particular, for domain walls the corpuscular corrections appear to be of order 
$\sqrt{\hbar g^2} $.

  \item  The corpuscular effects can break supersymmetry. 
  
  \end{itemize}
  
 Our key starting point rest on modelling the soliton as a coherent quantum state $|sol\rangle$ defined on the Fock space of corpuscles. These constituent quanta are  represented in this space in terms of 
 creation and annihilation operators,  $\hat{a}^+_{k}$ and $\hat{a}_{k}$,  satisfying the standard commutation 
 relations, $[\hat{a}_{k}, \hat{a}^+_{k'}]\, = \, \delta_{kk'}$ and $[\hat{a}^+_{k}, \hat{a}^+_{k'}]\, = 
 [\hat{a}_{k}, \hat{a}_{k'}]\, = \, 0$, where $k,k'$ are the momenta of the corpuscles.  The relation between 
 the corpuscles created by $\hat{a}^+_{k}$ and asymptotic one-particle states is not universal and 
 can be very complicated. However, very important conclusions can be reached without knowing the precise 
 dictionary. It suffices to know that the corpuscular description exists.  We  shall nevertheless identify the 
 explicit nature of the corpuscles on some examples.  
 
  The coherent quantum state, describing the soliton, is defined in terms of a set of data $N_k$.  The physical meaning of 
  $N_k$-s is that they set the average number of corpuscles of a given momentum $k$  in the quantum state 
  of the soliton.  That is, 
  \begin{equation}
    N_k \, = \, \langle sol| \hat{a}^+_{k}\hat{a}_{k}|sol \rangle \,.
  \label{Nkmeaning}
  \end{equation}
   The soliton is then described as the coherent state 
  that represents a superposition of the Fock basic vectors  $|n_{k=0} ....n_{k=\infty} \rangle  \, \equiv 
 \prod_k  \otimes  |n_k \rangle$ obtained by summing  over all possible distributions of 
  $n_k$-s, weighted by a factor determined by the fixed data set of $N_k$-s.  
  As we shall see this sum can be written in the form of a functional integral over  $n_k$-s, 
 \begin{equation}
|sol\rangle \, = \, \int \,  Dn_k \, {\rm e}^{S_{eff}} \, |n_k\rangle \, , 
\label{functional}
\end{equation}
 where, the effective corpuscular action is given by,  

 \begin{equation}
S_{eff}(n_k) \, = \, -  \frac{1}{2}\,\int dk \, \left( N_k \,  - \, n_k \ln N_k + \ln (n_k!)  \right) \, .
\label{actionintro}
\end{equation}
 This form immediately reveals the underlying nature of corpuscular corrections. 
 The notion of the classical saddle point is replaced by the most probable distribution
  $n_k^{(sd)}$ for given  data $N_k$, which, as we shall see, is given by 
  \begin{equation}
  n_k^{(sd)} \, = \, N_k \, - \, {1 \over 2} \, + \, O(1/N_k)\,.
 \label{SI}
 \end{equation}
   Corpuscular corrections come from the configurations 
 that account for the departures from $n_k^{(sd)}$  as well from the difference between 
 $n_k^{(sd)}$  and the classical saddle point  $n_k^{(class)} \,= \, N_k$. 
 
  It is now clear why these corrections cannot be captured by standard loop expansions and/or other semi-classical treatments.  All these standard corrections amount to correcting the data $N_k$ through, for example,  
 quantum renormalization of the masses and coupling constants of the theory.  But, the corpuscular corrections are not due to 
 corrections to $N_k$-s, but rather due to the contributions from the distributions that depart from 
the saddle point set by  $N_k$-s.  

Everything said so far only relies on the very existence of the corpuscular description and is not sensitive 
 to the particular nature of corpuscles.  It only assumes that the sensible data $N_k$ can be defined, 
 such that  in the limit $\hbar  = 0$,   the number of corpuscles  $N \equiv \int dk N_k = \infty$ and the  soliton physics is well-described by the classical saddle point, whenever such a description is available in the unresolved  classical theory. Quantum corpuscular corrections then scale like powers of ${1 \over \sqrt{N}}$, and consequently they vanish in the classical $\hbar=0$ limit.   
  
   The two interesting questions are the dictionary for  $N_k$ data-acquisition in terms of the classical soliton parameters 
   and the nature of the constituents.  Although, {\it a priory} there is no universal prescription,  in theories with spontaneous symmetry-breaking, in which the topological soliton has a well-defined classical limit,  
there is a clear way to define $N_k$-s, by identifying them with the coefficients of the Fourier  
expansion of the classical soliton solution,  schematically, 
\begin{equation}
   \phi_{sol}(z) \, = \, - \, \int \, dk \, ( e^{ikz}  i \sqrt{N_k}  \, + \, h.c.) \, , 
 \label{FexpN}
 \end{equation}
where $z$ stands for a space coordinate(s). 

 This way of mapping guarantees that the description has a correct classical limit.  It also allows for the identification of the constituents in the form of zero-frequency tachyons.  
  
    Such theories, when expanded about the unbroken symmetry phase (i.e. at the top of the so-called 
    ``mexican hat" potential), propagate tachyonic excitations 
  and are unstable.  The vacuum of the theories with spontaneous symmetry breaking is a vacuum in which tachyons are condensed.  However, the tachyons with momenta equal to the absolute value of their mass have zero frequencies and are {\it not}  unstable.  Such tachyons can form static configurations. 
  In this language, the topological solitons can be viewed as the  configurations in which some of the 
  zero-frequency tachyons are stabilized on top of the mexican hat potential by their momenta.  Thus, 
  in the corpuscular description the soliton can be viewed as a coherent state of such tachyons. 
  In this way of thinking about the soliton the notion of topological charge  emerges in the form of the tachyonic  momentum flow.   Incidentally, this makes clear why the topological charge cannot be carried by individual 
 particles that are asymptotic $S$-matrix states,  since such states cannot have tachyonic dispersion relations.   The topological charge is a collective phenomenon of quantum corpuscles 
 that are necessarily {\it confined}.  
 
  A class of theories in which the identification of the $N_k$-data is less straightforward  are the solitons which have no well-defined classical limit.  Such are the domain walls in $SU(N)$ super-Yang-Mills theories \cite{GiaS}, which for some time have been the source of the puzzle originally identified by Witten \cite{Witten3}.  It is clear that these domains walls cannot be described as solitons of any classical low energy field  theory. 
   Due to this, we cannot use the Fourier-transform method (\ref{FexpN}) for identifying the data $N_k$.   Nor we can identify tachyons as the constituent corpuscles.  However, as we shall briefly discuss, the theory offers
   natural candidates for the corpuscles in form of the UV-degrees of freedom, such as gauginos. 
  These UV degrees of freedom obey the same general property as tachyons: 1) They are confined and thus cannot be asymptotic states; 2) They conduct the chiral current across the wall and  this flow in the IR theory  translates as the topological charge.     
    
  Corpuscular effects can have important implications for supersymmetry breaking.
 The discussed way of visualizing the corpuscular physics makes it clear why we can expect that supersymmetry  is unable to cancel the corpuscular effects.  Supersymmetry is controlling the corrections to the data  $N_k$, by canceling some loop contributions  among bosons and fermions.  However, the corpuscular effects are not about  protecting the fixed data, but rather taking into the account the distributions that depart from it.

    Of course, one cannot exclude {\it a priory} that cancellations among the bosonic and fermionic excursions away from the saddle point could take place.  The main reason 
    why such cancellations do not happen is that fermions, due to Pauli exclusion principle, cannot 
    form a coherent state.  Therefore, the departures from the saddle point are mostly accounted 
    by variations of the bosonic distributions, whereas fermions contribute less.  
     
      We shall discuss the effect of supersymmetry breaking in the particular example of domain wall in Wess-Zumino theory, which is known 
  to preserve  half of the supersymmetry to all orders in perturbation theory.    We shall show that upon 
  the corpuscular resolution of the wall, also the second supercharge starts to act on the soliton state non-trivially and creates a second Goldstino.   In particular, we shall show that,
  \begin{equation}
   \hat{{\mathcal O}}_{BPS} |sol \rangle \, \neq \, 0 \,,  
  \label{BI}
  \end{equation} 
  where $\hat{{\mathcal O}}_{BPS}$ is the quantum operator counterpart of the classical BPS condition.  
 
  Corpuscular breaking of supersymmetry can have important implications for  explaining why we observe 
  no supersymmetry in Nature.  The idea that supersymmetry breaking is due to the fact that we live in a world-volume theory of a non-BPS topological defect goes back to \cite{GiaS2}. 
   Corpuscular breaking of supersymmetry can give a natural realization of this idea within string theory provided we  generalize the idea of corpuscular breaking to $D$-branes.  
   Such generalization would suggest that string theory has a built-in mechanism for supersymmetry breaking, 
   completely insensitive to the standard issues,  such as moduli stabilization.  Indeed, the corpuscular effects do not abolish either 
 Ramond-Ramond (RR) or topological charges, and therefore, they cannot destabilize branes that are stable in the classical limit. However, they make branes ``slightly" non-supersymmetric.   The estimated effect goes as $1/\sqrt{N}$, or 
   equivalently, as the string coupling $g_s$.  So the expected order parameter for supersymmetry breaking 
  would be $g_s$ times the brane tension scaling as $1/g_s$. Thus, the predicted supersymmetry breaking 
  scale is order one in string mass units.  Needless to say, the realization of this scenario would have very important phenomenological consequences. 
  
   The corpuscular way of thinking about non-perturbative objects opens up many obvious questions.  
  A natural direction would be the generalization of this way of thinking to Euclidean field configurations, such as instantons.
   Another question is whether there exist BPS configurations, such as extremal black holes or AdS spaces, that  are protected by supersymmetry against the corpuscular effects.

\section{Corpuscular Structure}

 Before formulating the effective Lagrangian approach to the corpuscular theory, we shall illustrate our philosophy 
 on a simple supersymmetric model.  Consider the classical Wess-Zumino model with the following superpotential,  
\begin{equation}
W \, = \, \Phi {1\over (L^2g)} \, - \, {g \over 3} \Phi^3 \,. 
\label{wclass}
\end{equation}
 Since we do not set $\hbar$ to one,  the dimensionality of  the chiral superfield 
 $\Phi$ (which is the same as dimensionality of its scalar component $\phi$) is 
 $[\Phi] \, = \, \sqrt{{mass \over length}}$. 
 The parameter $L$ has dimensionality of length and $g^2$ has dimensionality, $[g^2]\, = \, (mass\times length)^{-1}$. 
  
   This theory possesses two degenerate vacua with vacuum expectation values (VEVs) $\langle \phi \rangle \, \pm \, {1 \over Lg}$.   Both vacua are supersymmetric 
   with vanishing  VEV of the $F$-term.

   Because of the non-trivial vacuum topology, the theory also admits stable wall configurations 
   across which the classical field $\phi$ interpolates between the two VEVs.    
  These domain walls saturate the BPS bound \cite{GiaS}  and thus satisfy, 
\begin{equation}
\partial_z \phi_{sol}^* \, = \, \pm \left ({1\over (L^2g)} \, - \, g \phi^2 \right) \,.  
\label{BPSclass} 
\end{equation}
Taking parameters $L$ and $g$ real and positive, we have the real solution 
  \begin{equation}
 \phi_{sol}(z) \, = \, \pm \, {1 \over Lg}  {\rm tanh}  (z/L) \, ,   
 \label{kink}
 \end{equation}
 where $z$ is the coordinate perpendicular to the wall.  These are the so-called kink (antikink) solutions.   
 In the  classical theory in which so far we are working, each of these solutions is annihilated by one-half of the supercharges $Q_{\pm}$, corresponding to the sign of the eigenvalue of the $\gamma_z$-matrix, 
 $\gamma_z Q_{\pm} \, = \, \pm Q_{\pm}$.  
 
    In what follows we shall claim that corpuscular corrections to the kink spontaneously break 
  also the second half of supersymmetry.   In order to prepare a basis for understanding this effect, 
  we shall first  clearly identify the source of BPS-saturation in the classical theory.  The first thing is to understand the relation between the energy of the kink and its topological charge and to take into account this relation in the supersymmetry algebra. 
  
   The energy (per unit $xy$ area) of a $z$-dependent static classical configuration $\phi(z)$ is given by 
 \begin{equation}
   \rho \, = \, \int_{-\infty}^{+\infty} dz \, \left (K |d_z\phi|^2 \, + \, K^{-1} |\partial W/\partial \phi|^2 \right )  \, ,
  \label{tension}
 \end{equation} 
 where $K$ is the positive definite K\"ahler metric.  This quantity can be rewritten 
 as a perfect  square plus a  difference between the boundary values of the  superpotential,   
  \begin{equation}
   \rho \, = \, \int_{-\infty}^{+\infty} dz \,  |\sqrt{K} d_z\phi^* \, \pm \, \sqrt{K}^{-1} \partial W/\partial \phi|^2  \, \mp \,  (W(+\infty) \, - \, W(-\infty)).  
  \label{tensionB}
 \end{equation} 
 The quantity $T \, \equiv \, W(+\infty) \, - \, W(-\infty)$ is the topological charge.  
 This quantity is both holomorphic  as well as defined in terms of the VEVs. Due to this it is 
 neither subject to ordinary perturbative corrections, nor to the corpuscular corrections in 
 our sense.   Thus,  $T$ is fully determined by the classical parameters of the superpotential and 
 we do not need to worry about correcting it.   So let us focus our attention on the  perfect-square term in
 (\ref{tensionB}).  
 It is obvious that any wall configuration $\phi(z)$ that makes this term to vanish will correspond to a minimal 
 energy configuration in a given topological class, with energy being exactly equal to  $T$. 
 Such  classical configurations saturate the BPS bound. 
  Although K\"ahler is subject to usual perturbative renormalization, such corrections can be re-absorbed into the renormalization of the parameters of the theory and therefore they do not violate BPS-saturation. Since the corrections that we shall talk about are fundamentally different in nature, from now on and without loss of generality we shall set 
 $K \, = \, 1$.  
 
 One can immediately see that classical BPS configurations preserve half of the supersymmetry. 
 For this note that the algebra reads \cite{GiaS}, 
 \begin{equation}
     \big\{ Q,Q \big\} \, = \, {1\over 2} \gamma_{\mu} \gamma_0 \,  P^{\mu} \, + \, {1\over 2} \gamma_0 \gamma_5 \sigma_{\mu\nu} J^{\mu\nu}
   \label{algebraT} 
   \end{equation}
where $\sigma_{\mu\nu} \equiv {1 \over 2} [\gamma_{\mu}\gamma_{nu}]$ and  $J^{\mu\nu} \, \equiv \, \int d^3x \epsilon_{\mu\nu\beta 0}\partial^{\beta} W$ is the central charge of the
${\mathcal N} \, = \, 1$ superalgebra, which on the wall configuration is simply equal to the topological charge.  On the wall configuration  the algebra takes the form 
 \begin{equation}
     \big\{ Q,Q \big\} \, = \, {1 \over 2} \int dxdy \left ( \rho\, -  \, \gamma_z T \right ) \, .
   \label{algebraonwall} 
   \end{equation}
It is obvious that BPS-saturated states  $\rho\, = \pm T $,  will be annihilated 
by the supercharges that satisfy $ \gamma_z Q_{\pm} \, = \, \pm\,  Q_{\pm}$. 

   Correspondingly, the supercharge $Q_{\pm}$ will not lead to the fermion transformation, 
$\delta \psi \, = \, ( \partial_z \gamma_z \phi \, \pm \, F )\epsilon_{\pm} =  0$, where 
$ \gamma_z \epsilon_{\pm} \, = \, \pm\,  \epsilon_{\pm}$ is the supersymmetry transformation parameter. 
 The remaining supercharge $Q_{\mp}$ will create instead a Goldstone fermion, 
$\delta \psi_{Gold}  \, = \, 2F\epsilon_{\pm}$.   Notice that the $z$-dependent profile for the Goldstone 
fermion is identical to that of a scalar Goldstone mode  $\delta \phi(z) \, = \, d_z \phi(z)$. 
In other words,  the Goldstone fermion and a scalar form a supermultiplet under the  unbroken half supersymmetry. 

Notice also that by the BPS condition, the norms of both  Goldstone particles  $\int dz |F|^2$ are equal 
to the topological charge $T \, = \, {4 \over 3 L^3g^2} $.  Hence the decay constant of a canonically normalized Goldstino  is also set by the topological charge, $f_{Gold} \, =\, \sqrt{T}$.

  Obviously, for a wall to be a BPS-saturated object, the perfect-square term in (\ref{tensionB}) must be identically zero  for any value of $z$.  As was said above,  for the classical field $\phi(z)$ this condition can be satisfied even after all the perturbative quantum renormalizations of the parameters have been taken into account.\footnote{The fact that perturbative quantum renormalization of the  K\"ahler metric cannot violate the classical BPS condition,  can be understood in simple terms, for example, by applying an argument similar to the one by Witten \cite{Witten?} about the impossibility of perturbative supersymmetry breaking.  The above theory classically contains a single massless Goldstone fermion, whereas all other fermions are separated by a mass-gap.  In order for the perturbative corrections to spontaneously break a second half of supersymmetry, the second fermion must become massless and assume the role of Goldstino. This is impossible at any order in perturbation theory, since corrections to the tree-level fermion masses must be suppressed by powers of the weak coupling.  This argument is however not applicable to the non-perturbative corpuscular effects that will be discussed below.}

   The main question we are interested in is what happens once the classical wall field 
    is resolved into quantum corpuscles.  Such a resolution means that the entity 
    ${\mathcal O}_{BPS} \equiv d_z\phi \, \pm \, F$ has to be promoted into an operator, $\hat{{\mathcal O}}_{BPS}$, 
    acting on some quantum state 
    $|sol\rangle$, 
 which describes the kink in the Hilbert space.   The state $|sol\rangle$ must carry the information 
 about the corpuscular structure of the kink in such a way that the classical expression 
 is recovered as the $\hbar \, = \,0$ limit of the expectation value.     
 
  Notice, that simply replacing $\phi$ in the energy functional by its quantum counterpart quantized around the vacuum, will not take us far. Indeed,  the quantum constituents of the kink are interaction eigenstates and do not obey the dispersion relation of the free particles. So we need to identify a proper translation manual into the corpuscular language. 
  We shall try to do it by identifying the wave-modes of the classical kink field with expectation values 
  of some creation and annihilation operators.    
  
    Once this is done we shall discover that the
  achievement of the BPS equation in terms of expectation values, does not guarantee the 
  annihilation of the kink quantum state by the supercharge. 
  
     In order to see this, let us start resolving the kink into corpuscles.  For that we  consider the  Fourier  expansion 
 of the kink solution into  wave-modes,  
 \begin{equation}
   \phi_{sol}(z) \, = \, - \, \int \, dk \, ( e^{ikz}  i a_k \, + \, h.c.) \, . 
 \label{Fexpansion}
 \end{equation}
 The  expansion coefficients $a_k$ are just $c$-numbers given by 
  \begin{equation}
 a_{k} \, =  \, - \,{1 \over 4 g} \,  csch \left( {\pi  \over 2}kL \right ) \, .
 \label{kinkmodes} 
 \end{equation}  

  The corpuscular interpretation of these 
 $c$-numbers is that they represent the expectation values of the creation and annihilation 
 operators $\hat{a}_k^+, \hat{a}_k$ of the constituents taken over the quantum state of the kink,
 \begin{equation}
a_k \, = \,  \sqrt{\hbar}\,  \langle sol| \hat{a}_k |sol\rangle \,  .
\label{connection}
\end{equation} 
 The appearance of $\sqrt{\hbar}$ is crucial and this is where the information about the  quantum corpuscular structure enters.
 In order to have the correct classical limit any corpuscular resolution of the $c$-number Fourier coefficients 
 in terms of number operators must inevitably contain powers of $\hbar$.  This is also clear from the 
 dimensionality of $a_k$-s which is just $\sqrt{\hbar}$.   In other words, since $\hat{a}^+\hat{a}$ are interpreted as occupation number operators, the expectation values of $\hat{a}^+$ and $\hat{a}$ 
 must  scale as $\propto \, {1 \over \sqrt{\hbar} g} \,  \rightarrow \infty$ in the classical limit in which  
$\hbar \,  \rightarrow \, 0$ and $g$ is kept finite.

We shall model the state $|sol\rangle$ as a coherent state which represents a direct tensor product  
 \begin{equation}
 |sol\rangle  \, = \, \prod_k \, {\otimes} \, |N_k\rangle_{coh} \, 
 \label{solquant} 
 \end{equation}
 over coherent states  per each momentum $k$, 
   \begin{equation}
 |N_k\rangle_{coh} \, = {\rm e}^{-{N_{k} \over 2}} \, \sum_0^{\infty} \, {N_{k}^{{n_{k}\over 2}} \over \sqrt{n_{k}!}}  |n_{k}\rangle \,.  
 \label{coherent}
 \end{equation}
 By construction, the coherent state satisfies $\hat{a}_k |N_k\rangle_{coh} \, = \, \sqrt{N_k} |N_k \rangle_{coh}$. 
It is then obvious that 
$a_k \, = \, \sqrt{\hbar N_k}$.  The kink soliton is reproduced as the classical limit  for the distribution of $N_k$-s given by (\ref{kinkmodes}). \footnote{Notice that for $a_k$ given by  (\ref{kinkmodes}) 
the quantity $ \int_{-\infty}^{+\infty}  dk \,  N_k $ is divergent. This divergence is an artifact of the VEV of $\phi$ being a non-zero constant at infinity.  This can be easily understood by noticing that 
$ \int_{-\infty}^{+\infty}  dk \,  N_k  \, = \,  {1\over 8\pi \hbar }  \int_{-\infty}^{+\infty} dz \phi_{sol}^2(z)$.  
The finite quantity $N$ playing the role of  the integrated number of corpuscles is obtained by subtracting the  contribution of the constant, 
$ N \, \equiv \, \int_{-\infty}^{+\infty}  d(kL) \,  N_k'  \, = \,  - \, {L\over 8\pi \hbar}  \int_{-\infty}^{+\infty} dz ( \phi_{sol}^2(z) \, -  1/g^2L^2)$,  where $N_k' \, = \,  {1 \over 16\pi} \, (kL) \, {\rm csch} (\pi kL/2)$. 
The need for this subtraction only appears in the overall normalization of the probability weight functions. Alternatively, we can consider only the ratios of the probabilities.}

  Once we have constructed the kink state,  we define a quantum field operator $\hat{\phi}_{sol}$
  \begin{equation}
   \hat{\phi}_{sol}(z) \, = \, - \, i \, \int \, dk \, ( e^{ikz}   \hat{a}_k \, - \,  e^{-ikz}   \hat{a}_k^+) \,,  
 \label{Fieldoperator}
 \end{equation}
 for which 
 $\hat{a}_k^+$ and $\hat{a}_k$ are the creation and annihilation operators of the kink corpuscles. 
 For this quantum field the kink quantum state is a coherent state.  These creation and annihilation operators should not be confused with the analogous creation and 
 annihilation operators for the free quantum field $\phi$ quantized around any of the two classical vacua.  
 This is already obvious from the fact that the states created (or destroyed) by these operators satisfy very different dispersion relation from ordinary free particles. For instance, they have finite de Broglie wave-lengths but zero frequencies.  In general the relation between 
 $\hat{a}_k^+, \hat{a}_k$ and free field operators can be extremely complicated. However, 
in order to see the departure from BPS, the explicit connection with the free-field operator 
is not needed. It is enough to know that the quantum soliton state in terms of  $\hat{a}_k^+$ and $\hat{a}_k$ 
exists  at least in some corners of parameter and coordinate spaces.  
 In the next section, we shall identify such corners, in which the connections between the corpuscles  
 and the free-field quanta can be traced. 

   Before doing this,  let us focus on how the corpuscular structure causes departures from the BPS condition.  For definiteness, we shall consider the kink solution, corresponding to the choices of the plus signs in equations (\ref{BPSclass}) and (\ref{kink}),  and 
adopt the following terminology.  We shall denote by $\hat{{\mathcal O}}_{BPS}$ the operator that classically annihilates this state and corresponds to the 
unbroken supersymmetry. We shall refer  
to it as the BPS operator.  The parity-conjugated operator, that corresponds to the classically-broken 
supersymmetry, shall be denoted by  $\hat{{\mathcal O}}_{\overline{BPS}}$ and be called the 
anti-BPS operator. 
   
     In order to detect the corpuscular departures from the BPS bound, let us act on 
  the quantum state of the kink (\ref{solquant}) with  the BPS operator,    
    \begin{equation}
\hat{{\mathcal O}}_{BPS}  |sol\rangle \ = \, 
\left( \partial_z \hat{\phi}_{sol}^* \,-  \, {1\over (L^2g)} \, + \, g \hat{\phi}_{sol}^2 \right) \, |sol\rangle  \, .   
\label{BPSclassQ} 
\end{equation}
  This state vector,  represented as a superposition over a complete set of  the Fock space basis vectors,  $|n_{k =0} ... n_{k=\infty} \rangle  \, \equiv  \, \prod_k \, {\otimes} \, |n_k\rangle$, 
  has the following structure, 
  \begin{equation}
   \hat{{\mathcal O}}_{BPS}  |sol\rangle \ 
= \, \sum_{n_{k =0} ... n_{k=\infty}}   \, {\mathcal F}_{BPS} (z, n_k)  \, \times \sqrt{{\mathcal P} (n_k) } \, 
|n_{k =0} ... n_{k=\infty} \rangle  
\label{state1}  
\end{equation}
where,  
\begin{equation}
\sqrt{{\mathcal P}(n_k)} \equiv
\prod_k \, {\rm e}^{-{N_{k} \over 2}} \, {N_{k}^{{n_{k}\over 2}} \over \sqrt{n_{k}!}} \, ,
\label{matrixelement}
\end{equation}
 is the weight functional and 
 \begin{eqnarray}
 \label{coherentzero}
 & {\mathcal F}_{BPS} (z, n_k)   \, = \, \int dk  \Big\{k\sqrt{\hbar} \left( e^{ikz}  \sqrt{N_k}   \, + \, e^{-ikz} {n_k
  \over \sqrt{N_k} } \right )   \, - \, {1 \over L^2g} \delta(k) \, + \, \\ \nonumber 
& +\, g\hbar {1 \over L} \left ( e^{-i2kz} {n_k \over N_k}  \,  + \, 1\right ) \, \Big\} - \\ \nonumber 
 &- \, g\hbar \Big\{ \int dk \,\left( e^{ikz} \sqrt{N_k} \, - \,  e^{-ikz} {n_k \over \sqrt{N_k}} \right ) \Big\}^2 \,.
\end{eqnarray}
Here $N_k$-s are the fixed data, given by (\ref{kinkmodes}), fully determined by the classical limit of the given soliton, whereas $n_k$-s define all possible quantum distributions of the occupation numbers over the entire  Fock space.   Since,  each distribution $n$ is a function of $k$, what we are dealing with is a {\it functional sum}  
over functions $n_k$. This sum for large occupation numbers can be represented as the following functional integral,   
 \begin{equation}
\hat{{\mathcal O}}_{BPS}  |sol\rangle  \, = \, \int \,  Dn_k \, {\mathcal F}_{BPS} (z, n_k)  \, {\rm e}^{S_{eff}} \,  |n_{k =0} ... n_{k=\infty} \rangle \, ,  
\label{functional1}
\end{equation}
where,  $S_{eff} (n_k, N_k)$ is the {\it corpuscular effective action}  given by,  
\begin{equation}
S_{eff} (n_k; N_k)  \, \equiv  \, \ln {\mathcal P} (n(k);N(k)) \,,  
\label{Seffective1}
\end{equation}
or equivalently,  by (\ref{actionintro}).  This functional integral form is in full accordance with the  functional integral representation of the soliton given by (\ref{functional}).   

 We thus observe that  an  effective action  formulation in form of a functional integral is built-in  in the corpuscular 
picture of solitons described  as coherent states.   This language makes the nature of the corpuscular corrections very transparent in form of the departures of $n_k$-s from the most probable distribution,  $n_k^{(sd)}$.  
  For any fixed data $N_k$, the extrema of the weight functional (\ref{matrixelement}) (equivalently of the effective action)
 are achieved for the most probable distribution(s),   $n_k^{(sd)}$, which replace the notion of classical saddle point in the corpuscular picture.   We shall therefore refer to $n_k^{(sd)}$ as the quantum or 
 corpuscular  saddle point. 
  The corpuscular corrections are due to the departure 
of this quantum saddle point distribution,  $n_k^{(sd)}$, from the would-be classical saddle point value, 
 $n_k^{(class)} \, = \, N_k$, as well as due to the contribution of all other possible distributions $n_k \, \neq \, N_k$.
  
   We shall come back 
to the detailed discussion of the functional integral formulation in section 5.  However, we first wish to 
prepare the basis and complete the discussion in the language of probability distributions, without 
explicitly using the functional integral form.

  In order to see that the state (\ref{state1}) is non-zero,  we project it on an arbitrary number-eigenstate vector  
  $\langle n_{k =0} ... n_{k=\infty}|$. 
 The result of this projection is the removal of the sum,   
\begin{equation}
 \langle n_{k =0} ... n_{k=\infty}| \hat{{\mathcal O}}_{BPS}  |sol\rangle \ 
= \,  {\mathcal F}_{BPS} (z, n_k)  \, \times \sqrt{{\mathcal P} (n_k) } \, .
\label{project}  
\end{equation}
Notice, that both quantities ${\mathcal P}(n_k) $ and  ${\mathcal F}_{BPS}(z, n_k)$ depend on the same continuous set of  occupation numbers 
$(n_{k=0}, .... n_{k=\infty})$. The set contains a continuum of entries although each $n_k$ takes a fixed integer value per distribution.  In order for the BPS matrix element to be identically  zero, 
the above equation must be zero for all the distributions $n_k$  for which
${\mathcal P}(n_k) $ is non vanishing.  In particular, squaring the matrix element and summing it over a complete set of 
$n_k$-s we recover the expectation value over the soliton state of the $\hat{{\mathcal O}}^+_{BPS} \hat{{\mathcal O}}_{BPS} $  operator,  
\begin{equation}
\langle sol|\hat{{\mathcal O}}^+_{BPS} \hat{{\mathcal O}}_{BPS}  |sol\rangle \, = \, \sum_{n_k-sets} \,  |{\mathcal F}_{BPS}(z, n_k)|^2 {\mathcal P} (n_k) \, . 
\label{BPSsquare}
\end{equation}
In order for this quantity to be identically zero, the quantity ${\mathcal F}_{BPS}(z, n_k)$ must vanish for all the  $n_k$-s for which the weight function ${\mathcal P}(n_k) $ is non-zero.  

 Let us estimate the spread of $n_k$-s around the maximal value of the weight function.
 As said above, the maximal value of ${\mathcal P}(n_k) $ is what replaces the notion of saddle point  in the corpuscular theory.  We now wish to understand how sharply this value is peaked in the space 
 of $n_k$-s and  what are the corrections coming from departures of the classical saddle point distribution. 
 
 In order to see this,  let us define 
$P(n) \equiv  {\rm e}^{-N} \, {N^n \over n!} $  which only depends on a single $n$, 
    so that ${\mathcal P}(n_k)  =  \prod_k \, P(n_k) $. 
Using Stirling formula 
 for large $n$, we can approximate  
 \begin{equation}
P(n) \equiv  {\rm e}^{-N} \, {N^n \over n!} \simeq 
  {\rm e}^{-N } (Ne/n)^n {1\over \sqrt{n}}
\label{approx}
\end{equation} 
  This  function is maximized at  $N/n \, =\, {\rm e}^{1/2n}$, or equivalently, 
 \begin{equation}
 n_{(sd)} \, = \,  N\left(1\,  - \, {1 \over 2N}  \, + \, O({1/N^2}) \right) \, .
 \label{saddleC}
 \end{equation} 
 The value $n_{(sd)}$  determines the 
 most probable corpuscular distribution, which  replaces the notion of classical saddle point.
 Notice, that  $n_{(sd)}$ approaches the classical saddle point value $N$ only in the infinite $N$ limit. 
 Thus, the saddle point itself is corrected by $1/N$ corpuscular effects.  

    At its maximum  ${\mathcal P}(n) \,  \simeq 
 {1 \over \sqrt{N}}$.  For large $N$ the second derivative at the maximum is, 
  \begin{equation}
{d^2 P(n) \over dn^2}|_{n=n_{sd}}   \, \simeq \, - \, {1 \over  N \sqrt{N}} \, , 
\label{approx}
\end{equation}
which indicates that the  saddle point is flat and the spread of values of $n$ for which $P(n)$ is close to its maximal value is approximately $\Delta  n \, \sim \, \sqrt{N}$.  For large departures  $P(n)$ drops exponentially.

    On the other hand, since $n$ can only assume integer values,  for $N \, \ll  \, 1$,   $P(n)$ is maximized for $n=0$, with  $P(0) \,  = \,  {\rm e}^{-N}$.

 Thus, for any given data $N_k$,  the quantity  ${\mathcal P}(n_k) $ is maximized for the
  following set of  $n_k$-s:
  {\it 1)}  For $  N_K \, \gg \, 1$     $n_k\, \simeq \, N_K$ 
    with the spread $\Delta  n_k \, \sim \, \sqrt{N_k}$;    {\it 2)} For $  N_K \, \ll \, 1$,    $n_k\, = \,0$ 
    with zero spread;  {\it 3)}  For   $N_K \,  \sim 1$  the maximum is still at $n_k=0$, but changing 
   $n_k$ by order one is causing fluctuations  of order $1/n! \sim 1$.
    
  Applying this knowledge  to the soliton data  $N_k$, the appropriately-normalized (see footnote 3)
  saddle point  value of weight functional can be estimated to be, 
 \begin{equation}
{\mathcal P}(n_k)_{sd}  \,  \sim \,  g\sqrt{\hbar} \, .
\label{weightintegrated} 
\end{equation}
  Thus, the weight function contributes with this value  over a region of the Fock space with spread 
$\Delta  n_k \, \sim \, \sqrt{N_k}$. 

 We are now ready to investigate the corpuscular corrections to the function  ${\mathcal F}_{BPS} (z, n_k)$ 
 in the neighborhood of the saddle point. For definiteness let us consider the value at $z=0$,     
 \begin{eqnarray}
 \label{f0}
 & {\mathcal F}_{BPS} (z=0, n_k)   \, = \, \int dk  \Big\{k\sqrt{\hbar} \left( \sqrt{N_k}   \, + \, {n_k
  \over \sqrt{N_k} } \right )   \, - \, {1 \over L^2g} \delta(k) \, + \, \\ \nonumber 
& +\, g\hbar {1 \over L} \left ({n_k \over N_k}  \,  + \, 1\right ) \, \Big\} - \\ \nonumber 
 &- \, g\hbar \Big\{ \int dk \,\left(\sqrt{N_k} \, - \,  {n_k \over \sqrt{N_k}} \right ) \Big\}^2 \,.
 \end{eqnarray}
 Let us evaluate it at the saddle point.    As we have seen,  for the saddle point distribution 
 $n_k$-s very closely follows the data, $n_k \, =  \, N_k \, - \, 1/2$,   for $N_k \, \gg \, 1$ and vanish  for $N_k < 1$.  Thus, 
 all the $n_k$-dependent  integrals in (\ref{f0}) get cutoff above certain $k_*$ defined 
 by the condition that $N_{k_*} \, = \, 1$. Thus, given the explicit form of the data, the value of $k_*$ can be estimated as $k_* \, \simeq \, {2 \over \pi L} \ln \left ({1 \over4\sqrt{\hbar} g}\right)$, up to $\hbar g^2$-corrections.    We must stress that 
 the cutoff scale $k_*$ is not uniquely defined and the distributions with values of $k_*$ of the same order give comparable contributions.  Hence, for estimating the saddle point value, 
 we can  insert in  (\ref{f0}) the distribution  $n_k^{(sd)} \, =  \, (N_k \, - \, 1/2) \theta(k_* \, - \, |k|)$, 
 where $ \theta(k_* \, - \, |k|)$ is the step-function that vanishes for $k \, > \,k_*$ and is equal to one otherwise.    
 After taking into account that  $2 \int dk  k\sqrt{\hbar}  \sqrt{N_k} \,  = \, {1 \over L^2g}$ 
 as well as that $\sqrt{N_k} \, = \, - \, \sqrt{N_k} $  and performing some 
 re-arrangement  the expression (\ref{f0}) can be brought to the following form 
\begin{eqnarray}
\label{HHH}
 & {\mathcal F}_{BPS} (z=0, n_k^{(sd)})   \, = \, - \, 2 \sqrt{\hbar} \int_{k_*}^{\infty} \,  dk  k \sqrt{N_k}   \, - 
 \sqrt{\hbar} \int_0^{k_*} \,  dk  { k \over  \sqrt{N_k}} \, +  \, \\ \nonumber 
& +\, g\hbar 2 \int_0^{k_*} \, dk \, -  g\hbar \int_0^{k_*} \,  dk  { 1\over  N_k} \,. 
\end{eqnarray} 
Integration over $k$ gives, 
\begin{eqnarray}
 & {\mathcal F}_{BPS} (z=0, n_k^{(sd)})   \, = \,   \sqrt{\hbar} {4 \over \pi^2} {1 \over L^2} 
 \left ( 3 \ln \left({1 \over 4g\sqrt{\hbar}} \right ) \, + \, 1 \right) \, \\ \nonumber 
 & +\, g\hbar {2 \over \pi} {1 \over L^2}  \left( 4\, \ln \left({1 \over 4 g\sqrt{\hbar}} \right ) \, - \, 
 1 \right ) \, \, + \, O(g^2) \,. 
\end{eqnarray} 
 Thus, for the saddle point distribution the  quantity   ${\mathcal F}_{BPS} (z=0, n_k)$  is of the order of 
$\sqrt{\hbar} \ln\left({1 \over g\sqrt{\hbar}}\right)$.  Thus, the BPS operator, that in the classical theory annihilates 
the soliton state, in corpuscular theory gives a non-zero result of the order $\frac{1}{\sqrt{N}} \ln (\sqrt{N})$. 
 
This effect has to be compared with the analogous contribution 
for the anti-BPS operator corresponding, in the classical theory, to the broken half of supersymmetry. 
The expression for the anti-BPS functional,  ${\mathcal F}_{\overline{BPS}} (z=0, n_k)$ is analogous to 
(\ref{f0}) except  the relative sign between the first and other terms is the opposite and therefore no longer cancels the 
leading contribution of the constant $1/gL^2$-term.  As a result, we have ${\mathcal F}_{\overline{BPS}} (z=0, n_k) \sim {1/gL^2}$, which is non-zero in the classical limit.  Squaring (\ref{project}) and taking into account 
(\ref{weightintegrated}), we get that the transition probability caused by the BPS operator in the corpuscular theory 
is, 
\begin{equation}
 |\langle n^{(sd)}_{k =0} ... n^{(sd)}_{k=\infty}| \hat{{\mathcal O}}_{BPS}  |sol\rangle|^2 \sim  g\sqrt{\hbar} 
 \ln^2\left({1 \over g\sqrt{\hbar}}\right) \, . 
\label{squareBPS}  
\end{equation}
 The analogous transition  probability for the anti-BPS operator is, 
 \begin{equation}
 |\langle n^{(sd)} _{k =0} ... n^{(sd)}_{k=\infty}| \hat{{\mathcal O}}_{\overline{BPS}}  |sol\rangle|^2 \sim  {1 \over g\sqrt{\hbar}} \, .
\label{squareantiBPS}  
\end{equation}
The ratio of the above two probabilities measures the relative probability of detection of  the two Goldstinos 
in the number eigenstate $|n^{(sd)} _{k =0} ... n^{(sd)}_{k=\infty} \rangle$ given by the quantum saddle point distribution, and is of the order of,
\begin{equation}
  {P_{Gold_{BPS}} \over P_{Gold_{\overline{BPS}}}}\, \sim \, \hbar g^2\ln^2(g \sqrt{\hbar}) \, .
  \label{ratio-gold}
  \end{equation}  
 Let us now estimate the spread of the ${\mathcal F}_{BPS}(z, n_k)$-functional in the space of distributions.  
 In order to do that let us take its functional derivative with respect to the  $n_k$-s at $z=0$,
  \begin{eqnarray}
  \label{funcder}
  & {\delta {\mathcal F}_{BPS}(z, n_k) \over \delta n_k} |_{z=0}   \, = \,  \left ( k\sqrt{\hbar} {1
  \over \sqrt{N_k} }  \, + \, g\hbar {1 \over N_k}  \right)  + \, \\ \nonumber 
 &+ 2g\hbar \, {1\over \sqrt{N_{k}}} \, \int dk' \,\left( \sqrt{N_{k'}} \, - \, {n_{k'} \over \sqrt{N_{k'}}} \right ) \, .
\end{eqnarray}
Taking for example the saddle point distribution $n_k^{(sd)}$ we get,  $ {\delta {\mathcal F}_{BPS}(z, n_k) \over \delta n_k} |_{z=0}   \, = \,  k\sqrt{\hbar} {1
  \over \sqrt{N_k} }  \, + \, g\hbar {1 \over N_k} $, which is non-zero and of order $ \sqrt{\hbar}{1
  \over \sqrt{N_k} }$.  
  This means that the quantity  ${\mathcal F}_{BPS}(z=0, n_k)$ will in general stay order $\sqrt{\hbar}$ 
  over a functional spread $\Delta n_k \, \sim \,  \sqrt{N_k}$.  Remembering that the spread of the flat region 
 around the maximum value of ${\mathcal P}(n_k)_{sd}$ is also $\Delta n_k \, \sim \,  \sqrt{N_k}$, we see that 
  ${\mathcal F}_{BPS}(z=0, n_k)$  is order $\sqrt{\hbar}$ for the entire flat neighborhood of the saddle point. 
   
   Therefore, we conclude that the quantum bosonic contribution at the corpuscular level violates the BPS bound by order $g\hbar$-effect and creates a second Goldstone fermion.     
  
      Let us now discuss the fermionic contribution.   Up to normal ordering the fermionic contribution 
  to the expectation value of the Hamiltonian over the state $|sol\rangle$ is coming from the 
  expectation value of the Yukawa interaction term 
  $\langle sol |  \int dz \hat{\phi}_{sol}  \bar{\hat{\psi}} \hat{\psi}  |sol\rangle$.  Since the soliton is by construction a purely bosonic state, the effect of the fermionic part on  this expectation value reduces to the creation and annihilation of a single fermion on top of the soliton state.  
The fermions that can contribute to such matrix elements are not necessarily  quanta of a free fermionic 
field $\psi$ quantized about the vacuum with spontaneously broken symmetry.  Just in the same way as the operator $\hat{\phi}$  creates and annihilates the kink corpuscles,  $\hat{\psi}$ creates and annihilates fermionic  degrees of freedom that are the interaction eigenstates relevant for the solitonic state. 
 However,  the way the operator $\hat{\psi}$ encodes information about the soliton is different 
 from the semi-classical treatment.  

   In the semi-classical treatment,  the information about the soliton is
 embedded in $\hat{\psi}$, by simply quantizing the fermionic field on the classical soliton background.  Correspondingly, the modes 
 contained in  the expansion of $\hat{\psi}$ directly contain the information about the background. For example, in such a treatment the  Goldstone fermion would appear as one of the modes in the expansion of $\psi$ in the following way, 
 \begin{equation}
 \hat{\psi}(z,x, y,t) \, = \, F(z) \hat{\psi}_{Gold}(x,y,t) \, +  \,  {\rm massive~modes}
   \label{semi-class-exp} 
 \end{equation}
 where $\hat{\psi}_{Gold}$  is a $2+1$ dimensional quantum fermi field. In this treatment, the entire information about the collective phenomenon is encoded in the $c$-number profile function $F(z)$, obtained by solving the classical BPS equation.  
 
  In the corpuscular treatment,  we delegate the entire information about the collective phenomenon 
  to the state vector $|sol\rangle$ and to the rules of action of the  corpuscular creation and annihilation operators.  For example, the Goldstone fermion in this language is a certain one-fermion state on top of the coherent state of quantum corpuscles.  
  
   With this approach the expectation value $\langle sol |  \int dz \hat{\phi}_{sol}  \bar{\hat{\psi}} \hat{\psi}  |sol\rangle$  effectively factorizes $\langle sol |  \int dz \hat{\phi}_{sol}  |sol\rangle  \int dk 
 \langle sol | \hat{b} \hat{b}^+ |sol\rangle$ and vanishes since 
 $ \langle sol |  \hat{\phi}_{sol}  |sol\rangle \,$ is an odd function.

  \section{The Identity of the Corpuscles} 
     
   In order to identify the nature of the constituent quanta we shall investigate the corpuscular structure 
   around the $z=0$ point and show that here the soliton quantum state maps into a coherent state of zero-frequency tachyons.  
   
     Classically at this point the $\phi^2$ term is not contributing into the 
   BPS equation, but as we have seen,  it does contribute quantum mechanically.  Nevertheless, 
   for the clarity of the physical picture, we shall ignore this technical complication and   
  proceed in the following way.  We shall map the theory at $z=0$ into a linear theory 
  of free-corpuscles, and see whether in this treatment  we qualitatively reproduce the linear effect  
  of the previously discussed quantum theory.  
        
  This linear mapping allows us to clearly identify the corpuscles.  
   The point $z=0$ corresponds to the point where the classical expectation value of the field 
   $\phi$ is zero.   Theory expanded around this point is a theory of a scalar tachyon $\phi$ with
   the negative mass-square $m^2 \, = \, - \, \hbar^2 L^{-2}$ and a massless fermionic partner $\psi$. 
  Of course, tachyons are  not asymptotic states.   A system originally placed at $\phi=0$,  
 is unstable with respect to tachyon condensation.  Occupation numbers of tachyons with 
 imaginary frequencies grow until the system relaxes into a true vacuum.  In this way the tachyonic 
 vacuum is unstable. 
  However, notice that tachyons with the momenta exceeding the absolute value of the mass,
  $\hbar k \, \geqslant \, |m|$,  
 are stable. 
  Of particular interest are the tachyons  with momenta  exactly saturating this bound, 
  $\hbar k \, = \, |m|$.   
 These tachyons are special since they have zero frequencies, but carry non-zero momentum. 
 Due to this property,  they are the natural constituents of static field configurations.  
 In other words, non-zero momentum prevents tachyons 
  from condensation and stabilizes them on the top of the mexican hat potential.  Topological defects that realize an unbroken symmetry phase in their core 
  can be viewed as the coherent states of  such zero-frequency tachyons.  The topological 
  charge of the defect then emerges as the momentum flow of such tachyonic corpuscles.

  Let us discuss how the above picture works for the Wess-Zumino model expanded about the point $\phi=0$.  The tachyons with momentum $k \, = \, L^{-1} $ carry zero frequencies and do not condense. 
       To the linear order in $z/L$, the classical kink solution can be approximated as 
 a monochromatic  tachyon wave of wavelength $L$,   which corresponds to  
 $a_k \, = \, \delta (k - L^{-1}) (1/2Lg)$.  With this configuration, it is obvious that classical BPS equation is satisfied  to the linear order in $z/L$.  In our corpuscular picture  the corresponding quantum state 
 is a coherent state of tachyons with a single momentum  $k = 1/L$
  \begin{equation}
 |sol\rangle \, =  \, |N\rangle_{coh} \, = {\rm e}^{-{N \over 2}} \, \sum_0^{\infty} \, {N^{{n\over 2}} \over \sqrt{n!}}  |n\rangle \,, 
 \label{Lcoherent}
\end{equation}
where $ \sqrt{N} \,  = \,  {1 \over \sqrt{\hbar} g} $.  Since only the momenta $k =1/L$ enters, we have dropped the index $k$. 

Taking the expectation values, 
\begin{equation}
 \langle sol| \hat{a}_k |sol\rangle  \, = \,  \langle sol| \hat{a}_k^{+} |sol\rangle^* \, = \, \delta (k - L^{-1}) {1 \over 2 L\sqrt{\hbar} g} \, , 
 \label{VEVs}
 \end{equation}
 and plugging it in the equations (\ref{Fexpansion}) and (\ref{BPSclass} ) we see that the 
 BPS equation is indeed satisfied to the linear order  in $z/L$. However, the quantum state
 $|sol\rangle$ is not annihilated by the BPS operator,   
\begin{equation}
\left (- \, \partial_z \hat{\phi} \, \pm \, {1\over (L^2g)}  \, \right) \, |sol\rangle \, \neq \, 0 .  
\label{BPSoperator} 
\end{equation}
Although the expectation values $\hat{a}$ and $\hat{a}^+$ over the coherent state satisfy 
(\ref{VEVs}), the states 
 $\hat{a}_k |sol\rangle$  and $\hat{a}_k^{+} |sol\rangle$ are in general different. 
 
 For example, projecting  (\ref{BPSoperator}) on an $n$-particle state of $k=1/L$ corpuscles $|n\rangle$, we get 

  \begin{equation}
 \langle n| (- \, \partial_z \hat{\phi} \, \pm \, {1\over (L^2g)}  \, ) \, |sol\rangle \, = \, \pm \, 
  {1\over 2}  \sqrt{P(n)}
   \sqrt{N} \left( {n \over N}  \, - \, 1 \, \right )     
 \label{projectn}
\end{equation}
where $P(n)$ is the weight function defined in 
(\ref{approx}), and we have used $\sqrt{N} \, = \, 1/(\hbar g^2)$. 
Thus, the corpuscular resolution is telling us that  even the supercharge $Q_+$ that was unbroken in the classical theory,  now acting on the kink quantum state 
 creates a second Goldstino state, whose wave-function  profile is proportional to
  $(- \, \partial_z \hat{\phi} \, + \, {1\over (L^2g)}  \, ) \, |sol\rangle  \, \equiv \, |Gold_+\rangle$. 
The probability of finding the Goldstino in one of the basis $n$-particle states  
is, 
 \begin{equation}
| \langle n||Gold_+\rangle|^2 \, = \,  {1\over 4}  P(n) 
  N \left( {n \over N}  \, - \, 1 \, \right )^2   \, . 
\label{+probability} 
\end{equation}
This probability is vanishing at $n=N$, corresponding to a classical saddle point, and instead is 
peaked at $n \, = \, N \, +  \Delta n$ where $\Delta n \, \sim \sqrt{N}$.  This makes sense, as the probability 
to detect the second Goldstino in the vicinity of the saddle point must vanish for $N = \infty$.

   In order to understand better the large-$N$ limit, it is instructive to compare 
   (\ref{+probability}) with an analogous probability  for the other Goldstino corresponding 
   to the supercharge $Q_-$ that is already broken in the classical theory. 
  The probability for detecting this Goldstino in an $n$-particle sate is, 
  \begin{equation}
| \langle n||Gold_-\rangle|^2 \, = \,  {1 \over 4} P(n) \,  N  \left (3 \, + \, {n \over N} \right)^2\,, 
\label{+probability} 
\end{equation}
which is peaked around the saddle point.
    
    Let us now see, how the mismatch between the central charge and the energy can be interpreted 
 in the corpuscular language.    
   We shall derive this result  by acting on the soliton state $ |sol\rangle $
    with the operator  $\big\{ Q,S_0 \big\} $, where $S_0$ is the supercurrent density  ($Q \, = \, \int d^3x 
    S_0$)  and using the relation of the supersymmetry algebra, 
   \begin{equation}
    \big\{ Q, S_{0} \big\} \, = \, {1\over 2} \gamma^{\nu} \gamma_{0} \,  T_{\nu0} \, .
   \label{algebra} 
   \end{equation}

   Notice that at this level the algebra includes no central extension. The central topological charge
  will emerge as a result of the tachyonic momentum flow in the corpuscular theory.   
   The energy-momentum operator  $T_{\mu 0}$ when acting on $ |sol\rangle $ is getting the following two contributions. 
   First, is the energy density coming from the constant $F$-term which contributes with the  amount  $T_{00}  |sol\rangle
  \,  = \, 1/L^4g^2 |sol\rangle $. The second is the momentum contribution from tachyons, which carry momentum $\hbar/L$ per tachyon, but zero energy.  
     The soliton, being a coherent state on a non-zero $F$-term vacuum  while being an energy density eigenstate is not a momentum eigenstate, but rather a 
   Gaussian distribution of momentum eigenstates $|n\rangle$, with average value $N\, = \, 1/{\hbar g^2}$.   
 Thus we get,   
 \begin{equation}
  \big\{ Q,S_{0} \big\}  |sol\rangle  \, = \, {1\over 2} (  \,  T_{00} \, - \, \gamma_z \gamma_0 T_{0z}) |sol\rangle  \, .  
   \label{actionqq}
   \end{equation}
  This gives, \footnote{ Notice that in order for the Dirac operator  to have the correct square dispersion relation the contribution 
 from the tachyon momenta in $P_z$ must enter with one extra power of $\gamma_0$.
 This removes $\gamma_0$ infront of $\gamma_z$ in the last term.} 
  \begin{equation}
  \big\{ Q,S_{0} \big\}  |sol\rangle  \, = \, {\hbar \over L^4}
  {1\over 2}  \, \sum_n\, \sqrt{P(n)}  
  \, N \, \left( {n \over N} \gamma_z  \, - \, 1 \, \right ) |n\rangle     
    \label{actionN}
   \end{equation}

 Notice that the expectation value $ \langle sol|\big\{ Q,Q \big\}  |sol\rangle$ is zero for the 
 supercharges that satisfy $\gamma_z Q \, = \, \pm Q$, for kink and anti-kink respectively. 
 This is in full agreement with the earlier statement that the BPS equations are satisfied for expectation values. 
  However, the quantum state $\big\{ Q,Q \big\}  |sol\rangle$ is non-zero.  Again this can be obviously seen  
  by projecting this state on any given number eigenstate $|n\rangle$ which gives (where we choose 
  the supercharge to be the eigenvalue of the $\gamma_z$ that in the classical limit would annihilate the BPS state), 
   \begin{equation}
 | \langle n | \big\{ Q,S_0 \big\}  |sol\rangle|^2  \, =  {\hbar^2 \over L^8} \,  
  {1\over 4}  \, P(n) \,   
  \, N^2 \, \left( {n \over N} \, - \, 1 \, \right )^2     
    \label{square}
   \end{equation}

Which essentially reproduces (\ref{+probability}) for the Goldstino detection.  Similarly, 
this probability vanishes at $n=N$, corresponding to the classical saddle point, and instead is 
peaked  at $n \, \simeq \, N \, + \, \sqrt{2N}$.

\section{ Supershift of the Soliton in the Tachyon Picture}

 In order to check the self-consistency of the tachyonic description, 
  let us try to understand the action of a supersymmetric generator on the kink state from the point of view 
  of its action on the tachyonic quanta.   As we said, at $z=0$ the kink quantum state can be approximated 
  as a monochromatic coherent state of tachyons of momenta $p \, = \, {\hbar \over L}$ built on a 
  vacuum state  $|F\rangle$ with a constant $F$-term energy density.   However, the supersymmetric charge $Q$ is a global entity.  Thus, the action of $Q$ on the monochromatic tachyon state on the 
 constant $F$-term vacuum, will give the correct result only when such an approximation  of the soliton 
 state is globally valid.  This is only true in the limit  $g = 0$, $L=\infty, ~L^2g =$ fixed. In this limit $N$ is infinite, and thus half of the 
 $Q$-s must annihilate the  soliton.   Let us verify how the tachyonic coherent state description 
 reproduces this requirement.   In order to keep in mind that we work with infinite  $N$ and
 constant $F$-term vacuum, we shall denote each tachyon number eigenstate entering in the coherent state 
 as $|n, F\rangle \, = \,{(a^+)^n \over \sqrt{n!}} |F\rangle $.
   Acting on  the soliton state 
  by $Q$ we get,  
   \begin{equation}
   Q |sol\rangle  \, = \, 
   \sum_n\, \sqrt{P(n)}   Q {(a^+)^n \over \sqrt{n!}} |F\rangle    
    \label{actionQsol}
   \end{equation}
Using the commutation relations  $[Q_{\alpha} ,a^+] = b_{\alpha} ^+$, and $[a^+, b^+] = 0$ where $b_{\alpha}^+$ is a Fermi creation operator, 
we get, 

\begin{eqnarray}
& Q_{\alpha} (a^+)^n |F \rangle \, = \, \sqrt{n} \, b^+_{\alpha}  |n-1\rangle  \, +  \,  (a^+)^n Q_{\alpha}  |F\rangle \, = \, \\ \nonumber 
 & ( \sqrt{n} |n-1\rangle \times  |1_{\alpha} \rangle \, + \, \sqrt{N} \gamma^{z}_{\alpha\beta} |n\rangle  \times |1_{\beta}' \rangle ) \, .
  \end{eqnarray}
  The state  $|1_{\beta}' \rangle$ is a Goldstino produced by the action of the supercharge on 
  the $F$-term vacuum.  The factor $\sqrt{N}$ accounts for the Goldstino decay constant.

 Summing over $n$ in a coherent state we get 
 \begin{equation}
Q_{\alpha} |sol \rangle \, =  \, \sqrt{N}  |sol\rangle \times  \left ( |1_{\alpha} \rangle \, + \,\gamma^{z}_{\alpha\beta}  |1_{\beta}' \rangle \right ) \, ,
 \end{equation}
which is zero for  $|1_{\alpha} \rangle = - \gamma^{z}_{\alpha\beta}  |1_{\beta}' \rangle$. Thus we have reproduced, from the 
corpuscular point of view, the annihilation of the soliton by one-half of supersymmetry  in the $N=\infty$ limit.   For finite $N$, in order to get a trustable result, we must replace the global $Q$ by a local supercharge density $S_0$.   This reproduces the action of the BPS 
operator which results into the creation of  a Goldstone fermion  with the probability distribution given by 
 (\ref{+probability}).

\section{Corpuscular Effective Lagrangian and Quantum Corpuscular Effects}
In this section we shall introduce a functional integral representation of the soliton quantum state. This functional integral, fully equivalent to our previous description, will make explicit, in the standard functional integral language, the corpuscular resolution of the soliton state as well as the origin of the quantum corpuscular effects.    In this section we shall set $\hbar \, = \, 1$ and  shall slightly change the notations 
in order to rewrite index-dependences $N_k$ and $n_k$ into the functional form $N(k)$ and $n(k)$.

As already explained, the classical soliton configuration is characterized by a distribution $N(k)$. Associated with this distribution the quantum soliton state is defined in terms of the {\it quantum corpuscular distributions} $n(k)$ as follows
\begin{equation}
|sol\rangle \, = \, \int \,  Dn(k) \, {\mathcal P}(n(k);N(k)) \, |n(k)\rangle \, , 
\end{equation}
where we have introduced a notation $|n(k)\rangle \, \equiv \, |n_{k=0}, ...n_{k=\infty}\rangle$ and 
 we integrate over corpuscular distributions $n(k)$ with the weight 
\begin{equation}
{\mathcal P} \, (n(k);N(k)) \, = \,  e^{-\frac{1}{2}\int dk N(k)} \prod_{k} \frac{N(k)^{\frac{n(k)}{2}}}{\sqrt{n(k)!}} \, .
\label{Peffective}
\end{equation}
Notice that ${\mathcal P} \, (n(k);N(k))$ is exactly the same function as appeared in   
(\ref{matrixelement}).  

On the basis of this representation we can define the {\it corpuscular} effective action \footnote{
Notice that $\int dk$ has to be understood as the integration over a dimensionless argument 
of the function $N(k)$, which for domain wall example is given by $kL$, so that the effective action is dimensionless. 
However, for compactness of the notations we will not display this dependence  explicitly.} 
\begin{equation}
S_{eff} (n(k))  \, \equiv  \, \ln {\mathcal P} (n(k);N(k))
\label{Seffective}
\end{equation}
that yields
\begin{equation}
S_{eff}(n(k)) \, = \, - \, \frac{1}{2}\int dk \,  N(k) \,  + \, \int dk  \, \frac{1}{2} \, (n(k) \ln N(k) - \ln (n(k)!)) \, .
\end{equation}
Using $n!= \Gamma(n+1)$ and Stirling's series we get
\begin{eqnarray}
& S_{eff}(n(k)) = -\frac{1}{2} N +\frac{1}{2} \int dk n(k) ( \ln N(k) \, - \,  \ln (n(k) \, + \, 1) \, + \, 1) \, - \\  \nonumber
& \, - \, \frac{1}{4}\int dk \ln (n(k) \,  +\, 1) \, , 
\end{eqnarray}
where we have ignored the contribution of the extra terms $\sum_{l=1}^{l=\infty} \int dk \frac{B_{2l}}{2l(2l-1) (n(k) +1)^{2l-1}}$ with $B_{2l}$ being  the Bernoulli numbers.

Once we have defined the effective action we can approximate the corpuscular functional integral by the saddle point approximation. The saddle point defined by $\frac{\delta S_{eff}}{\delta n(k)} =0$ produces in our case the equation
\begin{equation}
\ln N(k) \, - \, \ln (n(k) +1) + \frac{1}{2n} \, + \,  O(1/n^2) \,  = \, 0 \, , 
\end{equation}
i.e.  the saddle point is a distribution $n_{sd}(k)$ very close to the distribution $N(k)$ defining the classical configuration. In other words, the saddle point of the corpuscular functional integral reproduces the classical distribution $N(k)$. More precisely we get
\begin{equation}\label{saddle}
n_{sd}(k) \, \simeq \, N(k) \left (1- \frac{1}{2 N(k)} \right ) \, .
\end{equation}
This equation captures the corpuscular correction to the saddle point.  Thus, in the corpuscular 
picture not only we account for the corrections coming from the off-saddle-point distributions $n(k) \neq n(k)_{sd}$, but 
the saddle point itself is corrected relative to its classical value.

\subsection{Quantum Corpuscular Corrections to the BPS Condition}

Quantum corpuscular corrections to the BPS bound can be obtained by evaluating the action of 
the BPS operator on the state  $|sol\rangle$. 

It is easy to check that
\begin{equation}\label{one}
\hat{{\mathcal O}}_{BPS}(z=0) \, |sol\rangle \, = \, \int \, Dn(k) \,  {\mathcal F}_{BPS} (n(k)) \, {\rm e}^{S_{eff}(n(k))} |n(k)\rangle \,, 
\end{equation}
where ${\mathcal F}_{BPS} (n(k))$ is given by (\ref{f0}).  Therefore, we shall automatically recover 
all the results obtained in section 2.

The quantum corpuscular corrections to the BPS condition arise because the function ${\mathcal F}_{BPS} (n(k))$ is non-zero over the space of corpuscular configurations $n(k)$ permitted 
by the effective action and therefore departures from saddle point,  i.e. contributions from corpuscular configurations different from $n_{sd}(k)$, define non-vanishing quantum corrections to the BPS condition in full accordance with what we have obtained in previous sections.

In order to estimate the general strength of the  quantum corpuscular corrections we can define
\begin{equation}
n(k) \, = \, n_{sd}(k) \, + \, n_{q}(k) \,, 
\end{equation}
and expand the effective action up to quadratic order in $n_q$
\begin{equation}
\frac{\delta^2 S_{eff}}{\delta n \delta n} \, = \,  - \, \frac{1}{2n} \, + \,  \frac{1}{4n^2} \, .
\end{equation}
Therefore, the order of magnitude of the quantum corpuscular effects is determined by
\begin{equation}\label{gauss}
\int D(n_q) {\rm e}^{-A n_q^2} \, = \, \sqrt{{\pi \over A}} \,  
\end{equation}
with 
\begin{equation}
A \, = \,  \int dk \left (\frac{-1}{2n_{sd}(k)} + \frac{1}{4n_{sd}(k)^2} \right ) \,,  
\end{equation}
which implies that the leading effect goes as $ \sim \frac {1}{\sqrt{N}}$.

At this point the reader could wonder why the underlying supersymmetry is not canceling the quantum corrections appearing from the integration over $n_q$. These {\it corpuscular quantum fluctuations} appear only after resolving the solitonic state into purely bosonic distributions $n(k)$ of constituents. Concerning the corpuscular dynamics encoded in $S_{eff}(n(k))$ the input about the classical configuration is in $N(k)$. The standard quantum fluctuations around the classical soliton induced by loop corrections define fluctuations of $N(k)$ that should be distinguished from the corpuscular fluctuations of $n(k)$. What the supersymmetry of the underlying Lagrangian does is to protect $N(k)$ from the perturbative quantum corrections.

  One can still think about some supersymmetry generalization of the corpuscular action $S_{eff}(n(k), f(k)^{\alpha})$ where $f(k)^{\alpha}$ represents distributions of fermionic corpuscles which for each $k$ and $\alpha$ can only take values $0$ or $1$.
     If we assume that under such generalization the notion of the classical soliton data $N_k$ is still 
  maintained in the sense of  expectation value, or equivalently, if we require that $|sol\rangle$ is a coherent state of bosons for each $k$, the possible modification of the state vector reduces to a direct multiplication by the fermionic distribution, 
  $\sum_{f_k^{\alpha}} \prod_{k, \alpha} |f_{k}^{\alpha}\rangle$. \footnote{In order to see this, let us 
  modify the coherent state for a given $k$ as $|N(k) \rangle \, = \, 
 {\rm e}^{-{N(k) \over 2}} \, \sum_{n(k)}  {N(k)^{{n(k)\over 2}} \over \sqrt{n(k)!}}   |n(k) \rangle \otimes 
 |f(k)_n^{\alpha}\rangle$. By taking the expectation value of $\hat{a}_k$ over this state, we get 
$\langle N(k) | \hat{a}_k|N(k) \rangle \, = \, \sqrt{N(k)} \sum_{n(k)}  {N(k)^{{n(k)\over 2}} \over \sqrt{n(k)!}} 
 \langle f(k)_{n+1} ^{\alpha} |f(k)_n^{\alpha}\rangle$. For this to be  equal to $\sqrt{N(k)}$,  the distributions $f(k)^{\alpha}$-s must be    
  $n$-independent, which reduces the modification to a direct product.}.

   Consequently, the fermionic and bosonic contributions to the functional integral factorize, 
  \begin{equation}
|sol\rangle \, = \, \prod_{\alpha}\left [\int Df_k^{\alpha}  |f_{k}^{\alpha}\rangle \right ] \otimes  \int \,  Dn_k \, {\rm e}^{S_{eff}
(N_k,n_k)} \, |n_k\rangle \, .  
\label{product}
\end{equation}
Given that the distributions of $n_k$ are arbitrary integers, whereas $f_k^{\alpha}$ are distributions of $0$-s and $1$-s, a generic cancellation in matrix elements cannot take place except  the usual vacuum normal-ordering-type cancellations, in which 
both $n_k$-s and $f_k^{\alpha}$-s take values $0$-s and $1$-s.

\subsection{A Comment on the Chern-Simons Meaning of the Corpuscular Effective Action}
The corpuscular effective action was defined using a representation of the soliton as a tensor product of coherent states for the tachyonic quanta. 
In terms of $N(k)$ and $n(k)$ this {\it corpuscular effective action} is given by
\begin{eqnarray}
&\int dk L_{eff} \, = \, - \, \frac{1}{2} N +\int dk  \frac{1}{2} n(k) (\ln N(k) -\ln (n(k)+1) +1) \, - \,  \\ \nonumber
& \frac{1}{4}\ln (n(k)+1)
\end{eqnarray}
We can interpret this corpuscular effective action as a genus expansion where the last term is the genus one contribution and where higher genus contributions are given by the terms $\frac{B_{2l}}{2l(2l-1) (n(k) +1)^{2l-1}}$ that we have ignored until now. In order to see the potential meaning of this corpuscular action let us focus on the genus zero term
$\frac{1}{2} n(k) (\ln (\frac{ N(k)}{(n(k)+1)}) +1)$. If for each value $k$ we redefine
$n\equiv N_{cs}^2$ and $\frac{ N(k)}{(n(k)+1)} \equiv \frac{t_{cs}}{e^{5/2}}$ we get the genus zero ( in planar sense) contribution to the free energy of three-dimensional Chern Simons theory with 't Hooft coupling $t_{cs}$ and gauge group $U(N_{cs})$ \cite{Witten},\cite{GoVa}
\begin{equation}
F\, = \, \frac{N_{cs}^2}{2}\left ( \ln t_{cs} - \frac{3}{2} \right ) \, .
\end{equation}

Of course,  this connection could be just a formal identification  without any deeper meaning \footnote{ For instance, the typical genus one term $-1/{12} \ln N_{cs}$ is not appearing under this identification with the right coefficient. The formal reason for this difference is simply that in the corpuscular coherent state picture we use the Gamma function while in the Chern-Simons case this role is played by the Barnes function.}. However, if pushed a bit further, this connection could mean that the corpuscular resolution of the wall is naturally connected with a superposition of three-dimensional Chern-Simons theories and that in the monochromatic limit where only one momentum contributes they produce just  a Chern-Simons with gauge group $U(\sqrt{N})$ with $N$ being the number of quantum constituents (per unit area). 

Very likely many features of the corpuscular Lagrangian are quite universal. What changes when we consider domain walls whose origin is quantum mechanical, as it is the case for super Yang Mills domain walls \cite{GiaS}, is the nature of the constituents that, as we shall see, become the confined UV degrees of freedom.

\section{Super Yang-Mills Domain Walls: Some Remarks}
For ${\mathcal N} \, = \,1$ supersymmetric  $SU(N)$ Yang-Mills we have a very explicit knowledge of the vacuum manifold. The number
of vacua is $N$ in agreement with the value of $Tr(-1)^F$ and each vacuum is characterized by the VEV
of the gaugino condensate  $ \langle \bar{\lambda} \lambda \rangle \, $ \cite {NVSZ} corresponding to the non-perturbative breaking of the $U(1)$ R-symmetry.   Because only the $Z_N$-subgroup is anomaly free, 
the vacuum manifold  has a discrete  degeneracy. 

The topology of the vacuum manifold 
 unambiguously indicates that there must exist domain walls interpolating 
 between the neighboring vacua \cite{GiaS}. Moreover, since the topological charge enters as a central term of the SUSY algebra, the tension of domain walls is BPS bounded. The crucial difference between this type of domain walls and the ones of the Wess-Zumino model that we have discussed before lies in the fact that in the Yang-Mills case the order parameter is composite, with the scale of compositeness 
 being of the same order as the thickness of the wall.

 This crucial difference manifests in the following well known puzzle. If the BPS bound is saturated the wall tension scales, in the large $N$ limit, as $N$. On the other hand if we try to imagine these topological defects as classical solitons of the IR effective low energy field theory, at leading order in $1/N$, we should guess the tension scaling as $N^2$. This mismatch, originally identified by Witten \cite{Witten3}, seems to indicate that these domain walls cannot be described as classical IR configurations. This state of affairs creates an obvious query, namely: what is the appropriate description of these topological defects in terms of the UV degrees of freedom? This way to pose the question makes perfect sense once we assume -- as it is in fact the case for ${\mathcal N} \, = \, 1$ super Yang-Mills -- that the theory is by itself UV-complete. 
 
 What can we say about this problem on the basis of the corpuscular resolution worked out above? The situation is more complicated than in the Wess-Zumino case where we have used the classical wall configuration to obtain the data $N(k)$ in terms of which we then defined the corresponding quantum coherent state. In the Yang-Mills case, since we do not have a classical field configuration to start with, we cannot fix the quantum state using any Fourier data. However, we know what quantity is  playing  the role of the field $\hat \phi$ of the Wess-Zumino model. 
 
  The wall connects two chirally asymmetric 
  vacua of the IR theory.  The motion from one vacuum to the other, from the point of view of the UV theory, 
  is equivalent to performing a coordinate-dependent chiral transformation 
  $\lambda \rightarrow e^{i\theta(z)\gamma_5} \lambda$, where $z$ is the coordinate perpendicular to the wall.   Thus, from the point of view of the UV theory the wall is a quantum state with a chiral gaugino current
 $j_{\mu}^5 \, \equiv \, \bar{\lambda} \gamma_{\mu}\gamma^5 \lambda$,  
   flowing in $z$-direction. In other words, the analog of the quantum field $\hat \phi_{sol}$ that we have used in the corpuscular resolution of the Wess-Zumino wall,  near the core $z=0$ of the wall is determined by the longitudinal part of the chiral current understood as a quantum operator written in terms of the UV degrees of freedom of the theory.  That is, the tachyonic momentum flow across the wall in Wess-Zumino case is replaced by the chiral gaugino current. 
  
  This fact reveals what quanta are now playing the role of the tachyons of the Wess Zumino model, namely these are the UV degrees of freedom. The phenomena of confinement takes care of locking these degrees of freedom within the wall i.e. takes care of the topological stability. 
  
  Although the concrete characterization of the quantum wall state $|wall\rangle$ is beyond the scope of the present paper,  we can nevertheless say few words about it. 
  
  The data fixing the state are determined by the flow of chiral charge i.e. $N$. This fixes the effective Fock vacuum $|N\rangle$ on which to define the analog of the bosonic states $|n(k)\rangle$. This state $|N\rangle$ can be thought as a bound-state of $N$ gauginos. This sort of bound state from the purely energetic point of view is analogous to a baryon bound state of $N$ fundamental quarks and fixes the tension of the wall state $|wall\rangle$ to scale as $N$. The fact that the quarks are in the adjoint representation  is not affecting this energetics qualitatively, since  gauginos couple with the strength 
  $1/N$,  although it is obviously affecting the notion of $N$-ality that we use with quarks in the fundamental
  representation. The states $|n(k);N\rangle$ are defined in terms of gluons and the anomaly identity $\partial j_5 \, = \, Tr F\wedge F$ (where $F$ is the gluon field strength) should fix these states as well as the effective corpuscular action. We leave the details of this construction for a future publication.

\section{Corpuscular Resolution of D-branes}

A natural way to approach the solitonic nature of $D$-branes is within the frame of tachyon condensation and K-theory \cite{sen},\cite{WittenK}.
For type $IIB$ $D$-branes 
the starting point is a configuration of equal number $\hat N$ of $9$-branes and anti $9$-branes. After GSO projection the corresponding world-volume theory contains a $U(\hat N)\otimes U(\hat N)$ gauge theory and a tachyon field, $T$, transforming  as $(N,\bar N)$. After assuming that the tachyon condensation breaks the gauge symmetry to $U(\hat N)$, $D$-branes are identified with topological lumps characterized by the 
non-trivial homotopy groups of the corresponding vacuum manifold that is isomorphic to $U(\hat N)$. Denoting by $U$ the map from the space-time manifold into $U(\hat N)$, the corresponding RR $p$-forms can be roughly represented as $G_p \sim Tr( U^{-1}dU)^p$. Instead, for type $IIA$ we must start with a set of $\hat N$ $D_9$-branes \cite{horava}. In this case the world volume theory contains a $U(\hat N)$ gauge theory, a tachyon in the adjoint representation and two chiral fermions equally transforming in the adjoint representation.  In this setup the RR $p$-forms can be defined using the $U(\hat N)$ gauge bundle. Denoting  by $F$ the corresponding gauge field strength we get $G_p \, \sim \, Tr F^{p/2}$.

In the case of Type $IIA$ we have $D_8$-branes that are domain walls in ten dimensions. From the tachyon condensation point of view we must choose a tachyon potential of the type 
\begin{equation}
V \, = \, \frac{1}{g_s} Tr f(T) \,,
\end{equation}
for some function $f(T)$ and with $T$ transforming in the adjoint representation. Supersymmetry is recovered away from the domain wall due to the condensation of the tachyons.  

In what follows we shall sketch how to extend the corpuscular resolution we have developed for 
Wess-Zumino domain walls to $D$-branes understood as tachyonic solitons. For type $IIA$ the main data used in the tachyon condensation design of $D$-branes is the number $\hat N$ of filling branes. The tension of the state defined by this set of filling branes in string units is $\hat N\frac{1}{g_s}$. As it is customary we can assume that $g_s =1/\hat N$. The dictionary into the corpuscular language used for the Wess-Zumino model can be simply derived by identifying the heights of the tachyon potential at $T=0$. This leads to
\begin{equation}
\hat N^2 \, = \, \int N(k) \equiv N \,
\end{equation}
and $L\, =\, L_s$. Generically, for co-dimension larger than one defect,  the radial mode defining the modulus of the tachyon field will assume momentum $k$ in the orthogonal direction to the $D$-brane.
However, the  phase modes will have momenta in the angular direction in order to account for  the topologically-nontrivial winding numbers. 
 
 The corpuscular effects as we have described for the Wess-Zumino model scale as  $1/{\sqrt{N}}$ that under the present correspondence becomes,
\begin{equation}
\frac{1}{\hat N} \, = \, g_s \,.
\end{equation}
We thus expect that in the corpuscular treatment the departure from the BPS 
condition should be of order $g_s$ and correspondingly the magnitude of supersymmetry-breaking in the world-volume theory should be by factor of order $g_s$ suppressed relative to the brane tension. Since the latter scales as $1/g_s$, the expected effective scale of supersymmetry-breaking is  the string scale.

 \section{Phenomenological Implications of Supersymmetry Breaking} 
 
   Generalizations of corpuscular ideas about supersymmetry-breaking can have important implications   
for understanding why we do not observe supersymmetry at low energies.  The idea that we 
do not see supersymmetry because we live in the world-volume theory of a non-BPS brane goes back 
to \cite{GiaS2}.  However, realizations of this scenario within string theory  faces number of  well-known difficulties, such as the instability of non-BPS  branes.  The simplest non-BPS brane configurations, e.g., such as the 
$D$-brane-anti-$D$-brane pairs, are known to be unstable in the semi-classical treatment. 
  
   The approach that we are offering here suggest a completely different view of supersymmetry breaking
   within string theory.   Applying the corpuscular picture to  $D$-branes would reveal that string theory has a
   built-in mechanism  for supersymmetry breaking,  free of any stability issues. 
   Indeed, the corpuscular effects only create a small mismatch between the RR  charges and the tensions, but 
   of course cannot  jeopardize the RR  or  topological charges that stabilize the branes. 
  In this way,  any string theory background that  contains non-perturbative entities that can be subjected to  
  corpuscular resolution could violate supersymmetry. 
  
  Since the corpuscular effect of SUSY-breaking goes as ${1 \over \sqrt{N}}$, translated in terms of the string coupling 
  this would amount to  an order $g_s$ effect.   As explained above, applied to the $D$-brane backgrounds, the predicted scale of supersymmetry breaking is as high as the string scale!

\section{Some other general implications of the corpuscular picture} 

 The ideas displayed in this paper open up a new way of thinking about solitons and non-perturbative entities in general. In order to push the picture forward the obvious steps would be to apply our treatment to other saddle 
point solutions, both in spaces with Minkowski or Euclidean signatures, such as instantons.   
Although, the specific details will vary, we expect that the general points should remain very similar. 
 For example, all the theories that give rise to topological defects that are produced as the result of spontaneous symmetry breaking, 
 admit the description of the vacuum in form of a tachyon condensate and correspondingly the
 description of the hill-top vacuum as a tachyonic vacuum. Consequently to visualize the core structure of topological defects as a coherent state of tachyons is in principle possible.  

   The  presented picture also raises an interesting question on how the corpuscular treatment affects those theories that display soliton/particle dualities.

    \section*{Acknowledgements}

The work of G.D. was supported in part by Humboldt Foundation under Alexander von Humboldt Professorship,  by European Commission  under 
the ERC Advanced Grant 226371 and ERC Advanced Grant 339169 ``Selfcompletion'',   by TRR 33 \textquotedblleft The Dark
Universe\textquotedblright\   and  by the NSF grant PHY-0758032. 
The work of C.G. was supported in part by Humboldt Foundation and by Grants: FPA 2009-07908, CPAN (CSD2007-00042) and HEPHACOS P-ESP00346 and by the ERC Advanced Grant 339169 ``Selfcompletion'' .


\begin{thebibliography}{99}
\bibitem{GiaCearN}G. Dvali , C. Gomez, Black Hole's Quantum N-Portrait,  
 Fortsch.Phys. 61 (2013) 742-767,  [arXiv:1112.3359]; 
 Quantum Compositeness of Gravity: Black Holes, AdS and Inflation 
JCAP01(2014)023,  [arXiv:1312.4795]. 

\bibitem{GiaS} G. Dvali and M. Shifman, Domain walls in strongly coupled theories. Phys.Lett. B396 (1997) 64-69, Erratum-ibid. B407 (1997) 452, [hep-th/9612128].  
 
 
\bibitem{Witten3} E. Witten, Branes and the dynamics of QCD. Nucl.Phys. B507 (1997) 658-690,  [hep-th/9706109]. 

\bibitem{GiaS2}G.R. Dvali, M. A. Shifman, Dynamical compactification as a mechanism of spontaneous supersymmetry breaking, 
 Nucl.Phys. B504 (1997) 127-146, [hep-th/9611213]
   

\bibitem{Witten?} 
E. Witten, Constraints on Supersymmetry Breaking, 
Nucl.Phys. B202 (1982) 253. 

\bibitem{Witten}E. Witten, Chern-Simons gauge theory as a string theory, 
 Prog.Math. 133 (1995) 637-678,  [hep-th/9207094].  
 
 \bibitem{GoVa} R. Gopakumar and C. Vafa, On the gauge theory / geometry correspondence,  
Adv.Theor.Math.Phys. 3 (1999) 1415-1443, [hep-th/9811131].   

\bibitem{NVSZ} 	V.A. Novikov, M. Shifman, A.I. Vainshtein and V. Zakharov, Instanton Effects in Supersymmetric Theories,  Nucl.Phys. B229 (1983) 407.

\bibitem{sen} A. Sen, Tachyon condensation on the brane anti-brane system . JHEP 9808 (1998) 012, 
 [hep-th/9805170]. 
\bibitem{WittenK}E. Witten, D-branes and K theory,  JHEP 9812 (1998) 019, [hep-th/9810188]. 
\bibitem{horava} P. Horava, Type IIA D-branes, K theory, and matrix theory,  Adv.Theor.Math.Phys. 2 (1999) 1373-1404,  [hep-th/9812135].  
 
\end{thebibliography}
\end{document}